\documentclass[aps,12pt,showpacs]{revtex4}

\usepackage[english]{babel}
\usepackage{amsmath,amssymb,amsbsy,amstext}
\usepackage[dvips]{graphicx}

\newcommand{\be}{\begin{eqnarray}}
\newcommand{\ee}{\end{eqnarray}}
\newcommand{\bdm}{\begin{displaymath}}
\newcommand{\edm}{\end{displaymath}}
\newcommand{\ba}{\begin{array}}
\newcommand{\ea}{\end{array}}
\newcommand{\ds}{\displaystyle}

\newcommand{\pa}[1]{\left(#1\right)}
\newcommand{\paq}[1]{\left[#1\right]}

\newcommand{\tl}[1]{\tilde{#1}}

\begin{document}

\title{Sensitivity of a small matter-wave interferometer
to gravitational waves}
\author{Stefano Foffa \protect\( ^{(1)}\protect \), Alice Gasparini
\protect\( ^{(1)}\protect \), Michele Papucci \protect\( ^{(2)}\protect\)
and Riccardo Sturani \protect\( ^{(1)}\protect \)}
\affiliation{
(1) Department of Physics, University of Geneva, CH-1211 Geneva,
Switzerland\\
(2) Department of Physics, University of California, Berkeley and Theoretical
Physics Group, Ernest Orlando Lawrence Berkeley National Laboratory, Berkeley,
CA 94720, USA\\
\texttt{e-mail: papucci@berkeley.edu;}\\
\texttt{Stefano.Foffa,Alice.Gasparini,Riccardo.Sturani@physics.unige.ch}}
\begin{abstract}
We study the possibility of using matter wave interferometry
techniques to build a gravitational wave detector.
We derive the response function
and find that it contains a term proportional to the derivative
of the gravitational wave, a point which has been disputed recently.
We then study in detail the sensitivity that can be reached
by such a detector and find that, if it is operated near resonance, it can 
reach potentially interesting values in the high
frequency regime.
The correlation between two or more of such devices can further
improve the sensitivity for a stochastic signal.
\end{abstract}

\maketitle

\section{Introduction}

The use of a matter wave interferometer for the detection of
gravitational waves has been recently proposed in
\cite{Chiao:2003sa}.
Analogously to what happens with light interferometers, matter-wave ones
are meant to measure the phase shift induced by the passage of a gravitational
wave (GW); however, dealing with atoms rather than photons implies some
crucial differences.

For instance, what is measured is the effect of the GW on the speed, rather 
than on the position, of freely falling objects (which are the interfering 
atoms in atomic interferometers and the mirrors in the traditional light ones).

More than that, a matter wave interferometer is sensitive also to the time
derivative of the GW, $\dot{h}$, thus making it convenient to operate
it in the high frequency regime.

The analysis performed in \cite{Chiao:2003sa} contains actually
some flaws. To begin with, the movement of the various parts of the
interferometer has not been correctly taken into account for
frequencies above the mechanical resonance.
As a result, the actual sensitivity attains its maximum value at the 
resonance, dropping sharply above it.
This means also that the size of the device can not be too
large, otherwise the lowest mechanical resonance would be too low
and the sensitivity at that point just not good enough.

Secondly, as noticed in \cite{Roura:2004se},
the passage of the gravitational wave affects also the departure times
of the two interfering paths of the atomic beam, a fact that has
not been included in the original proposal.
Actually the authors of \cite{Roura:2004se} go further and state that,
once this effect is taken into account, the phase shift does not
depend anymore on $\dot{h}$, thus completely spoiling the
interesting high frequency behavior of the response function.
We claim that this statement is incorrect, because it
is based on the crucial assumption that the atoms'
trajectories of both arms have exactly fixed endpoints,
coincident with the detector location.
In other words, the authors of \cite{Roura:2004se} have analyzed only the case
in which the passage of the GW does not change at all the extrema of the
atoms' trajectories. 
Such constraint is not physically motivated as the typical
displacement induced by a GW on the atom position is many orders
of magnitude smaller than the transverse spread of the atomic
wave-packet, not to mention the detector acceptance, thus not preventing the 
atomic beams from interfering and being detected.

We indeed find that, once the fixed endpoint condition is relaxed,
a term proportional to $\dot{h}$ is still present in the expression
of the phase shift.

We then explore the potentiality of a GW detector
inspired on (an amended version of) the proposal
of \cite{Chiao:2003sa}, and we require it to be based on
{\em present technology}. 
Such a device is thus in principle immediately realizable at relatively
low cost and in several copies.

The paper is organized as follows: in the next section we reproduce
the relation between the dephasing measured at the interferometer and
the amplitude of the GW.
The discussion will be quite detailed, in order to emphasize the points of
contact and disagreement with the treatment given in \cite{Roura:2004se}.
In section \ref{noise} we then examine the principal sources of noise that can
limit the sensitivity of a typical state of the art interferometer
and we plot and discuss the relative sensitivity curve.
Measurement of static or quasi-static gravitational or inertial fields by 
means of matter interferometers is already a mature discipline:
see for instance the pioneering work of \cite{Colella:dq} and the recent
measurement of the Sagnac effect reported in \cite{sagnac}.
The extension of this discipline towards the detection of more rapidly
varying signals is not a trivial task and, as it is a crucial point
in our treatment, we will discuss in detail the main limitations to
a fine time resolution of the signal, and consequently to the detection of 
high frequency GW's.
In section \ref{corr} we will discuss the possibility of correlating
the output of two small atomic interferometers and we will compare the
obtained sensitivity with the actual bounds on stochastic GW signals.
Final comments and remarks will conclude the paper.

In all our equations we will write $\hbar$ explicitly (not to be confused with
the gravitational perturbation $h$), while keeping the usual notation for 
the speed units $c=1$.

\section{The response function}\label{response}
Here we compute and discuss the response function for the GW detector based on 
atomic interferometry.
\subsection{General considerations}
The spacetime metric in presence of a GW propagating along the $\hat n$
direction can be parameterized in the TT gauge as
\be \label{geod}
{\rm d}s^2=-{\rm d}t^2+[\delta_{ij}+h_{ij}(t-\hat n\cdot \vec{x})]{\rm d}x^i 
{\rm d}x^j\,,
\ee
with $h^i_i=n^ih_{ij}=0$. This corresponds to the choice of a
free-falling reference frame.

We find convenient to work in a different gauge, 
related to the TT one by the coordinate transformation
\be
t\to t'=t\qquad x^i\to {x^i}'=x^i-\frac 12 h^i_jx^j\,,
\ee
which returns the metric 
\be \label{rigid}
{\rm d}s^2=-{\rm d}t^2-{\rm d}t{\rm d}x^i\dot h_{ij}x^j+
[\delta_{ij}+(\dot h_{ik}x^kn_j+\dot h_{jk}x^kn_i)/2]{\rm d}x^i {\rm
  d}x^j\, ,
\ee
where it is understood that the components $h_{ij}$ of the metric perturbation 
are the same as in the TT gauge.
In this coordinate frame the connection coefficients 
vanish at the origin of coordinates, making geodesic equations 
look like geodesic deviation equations. 
Moreover it can be read from eq.~(\ref{rigid}) that the GW does not
affect physical distances in the plane orthogonal to $\hat n$.
Because of this, in \cite{Roura:2004se} the coordinate frame
specified by (\ref{geod}) is named \emph{geodesic} and the (\ref{rigid}) 
\emph{rigid} frame.

In the latter, the explicit non-relativistic form (including the next to 
leading term in the velocity) of the action returning the geodesic 
equation at first order in $h_{ij}$
\be\label{geoeq}
\dot{V}^i=\frac{1}{2}\ddot h^i_j
X^j (1-2\hat n\cdot \vec V)-\frac 12\dot h_{jk}V^jV^kn^i
\ee
is then
\be \label{action}
S=\frac m2 \int \left(\vec{V}^2 -
\dot{h}_{ij}V^i X^j(1-\hat n\cdot \vec V)\right){\rm d}\tau\, ,
\ee
with $V^i\equiv\dot X^i$. 

Since in a typical interferometer the atom wave function is essentially peaked
around the two classically allowed trajectories (which will be called the
$(+)$ and $(-)$ paths henceforth), the phase in the final point can be 
calculated, for a given path, as:
\be\label{phase0}
\Phi(f)=\Phi(i)+\frac{1}{\hbar}S(i\rightarrow f)\, ,
\ee
where $S(i\rightarrow f)$ is the classical action from the initial to the final
point. 
Notice that in general one has $\Phi(i)^{(+)}\neq\Phi(i)^{(-)}$ since
the flight times along the two paths are not the same, because atoms
reaching the detector at the same final time, have entered the
interferometer at different initial times.

It is now useful, in order to calculate the action $S(i\rightarrow f)$, 
to introduce the splitting of the atoms position and velocity 
$\vec X,\vec V$ as $\vec X=\vec X_0+\delta \vec X$, 
$\vec{V}=\vec{V_0}+\vec{w}$, where $\vec X_0,\vec V_0$ are the
position and velocity in absence of the GW.

Taking also into account, as pointed out
in \cite{Roura:2004se}, the shifts $\delta t_{i,f}$ of the initial and final 
times $t_{i,f}$ of the trajectory the action reads as follows
\be \label{action1}
\renewcommand{\arraystretch}{1.4}
\ba{l}
\displaystyle
S\simeq S_0+m \int_{t_i+\delta t_i}^{t_f+\delta t_f}\paq{\vec{V_0}\cdot\vec{w}
-\frac{1}{2}\dot h_{ij}V_0^i X_0^j(1-\hat n\cdot\vec V_0)} {\rm d}\tau\simeq \\
\displaystyle
\quad\simeq S_0+m\int_{t_i}^{t_f} \paq{\vec V_0\cdot\vec w -
\frac 12\dot h_{ij}V^i_0X^j_0 (1-\hat n\cdot \vec V_0)}{\rm d}\tau+
\frac{m}{2} V_0^2(\delta t_f-\delta t_i)\,,
\ea
\renewcommand{\arraystretch}{1}
\ee
where $S_0$ is the value of the action computed over the unperturbed
trajectory.

Up to first order in the GW perturbation , the initial phase for each 
path is $\Phi(i)=-\frac{m}{2 \hbar} V_0^2 (t_i + \delta 
t_i)$.
This result allows us to rewrite eq. (\ref{phase0}) as
\be\label{phase2}
\frac{\hbar}{m} \Phi(f)= \frac{S_0}{m} - \frac{V_0^2}{2} t_i+ 
\int_{t_i}^{t_f}
\paq{\vec{V}_0\cdot\vec{w}-\frac 12\dot h_{ij}V^i_0X_0^j
(1-\hat n\cdot \vec{V}_0)}{\rm d}\tau+
V_0^2(\delta t_f-\delta t_i) -\frac{V_0^2}{2}\delta t_f\, .
\ee
We now notice that the physical distance between the initial (final)
point of the perturbed trajectory from the unperturbed one is
\be\label{DelX}
{\Delta\vec X}_{i,f}=\delta\vec{X}_{i,f}+\vec{V}_0\delta 
t_{i,f}\,,
\ee
and that 
$\delta\vec{ X}_f-\delta\vec{ X}_i=\int_{t_i}^{t_f}\vec{w}{\rm d}\tau$,
thus giving
\be\label{phase3}
\frac{\hbar}{m} \Phi(f)= \frac{S_0}{m} -
\frac{V_0^2}{2}\left(t_i+\delta t_f\right)-\frac12\int_{t_i}^{t_f} 
\!\!\!\dot h_{ij}V^i_0X_0^j
(1-\hat n\cdot \vec{V}_0)
{\rm d}\tau+\vec{V_0}\cdot({\Delta\vec X}_f-{\Delta\vec X}_i)\, .
\ee
The first two terms do not provide any information about the GW:
the first does not depend on $h$, while the second is
equal for both paths (since the atoms arrive at the detector simultaneously)
and thus cancels out when taking the phase difference.
We also notice that the term proportional to $\vec{w}$ has disappeared or,
better, it has been rewritten in disguise as part of the term in 
${\Delta\vec X}$.

The analysis performed so far is equivalent to the one done in
\cite{Roura:2004se}, whose authors however further elaborate on
eq.~(\ref{phase3}) by setting ${\Delta\vec X}_{i,f}=0$. This implies a
restriction only to such $h_{ij}$'s that do not change at all \emph{both}
the coordinates of the extrema of the atoms' trajectories. However, as stated 
in the introduction, any position change induced by the GW is much smaller
than the transverse spread of the atomic wave-packet, and than any detector 
acceptance.
This means that the interference and the detection take place also
for a GW affecting the endpoints of the atoms' trajectories,
and the condition imposed in \cite{Roura:2004se} does not describe a 
realistic experimental setup.

Clearly, by setting to zero the last term in eq.~(\ref{phase3}),
one loses any possible information on $\dot{h}$, since the only remaining term,
the next to the last, can be calculated by parts and shown not to contain any 
derivative of the GW. This is the ultimate reason of the claim contained in
\cite{Roura:2004se} about the absence of a term in $\dot{h}$ in the expression
of the phase shift.

It is useful to end this subsection by noticing that one
component of ${\Delta\vec X}_{i,f}$ can be expressed in terms
of the other by means of the direct substitution of $\delta t_{i,f}$ in
eq.~(\ref{DelX}), thus giving
\be\label{YofX}
\Delta Y_f -\Delta Y_i=\int_{t_i}^{t_f}
w^y{\rm d}\tau+\frac{V_0^y}{V_0^x}\left(\Delta X_f -\Delta X_i-
\int_{t_i}^{t_f} w^x{\rm d}\tau\right)
\,.
\ee

\subsection{The Mach-Zender configuration}
\begin{figure}
\centering
\includegraphics[width=.65\linewidth]{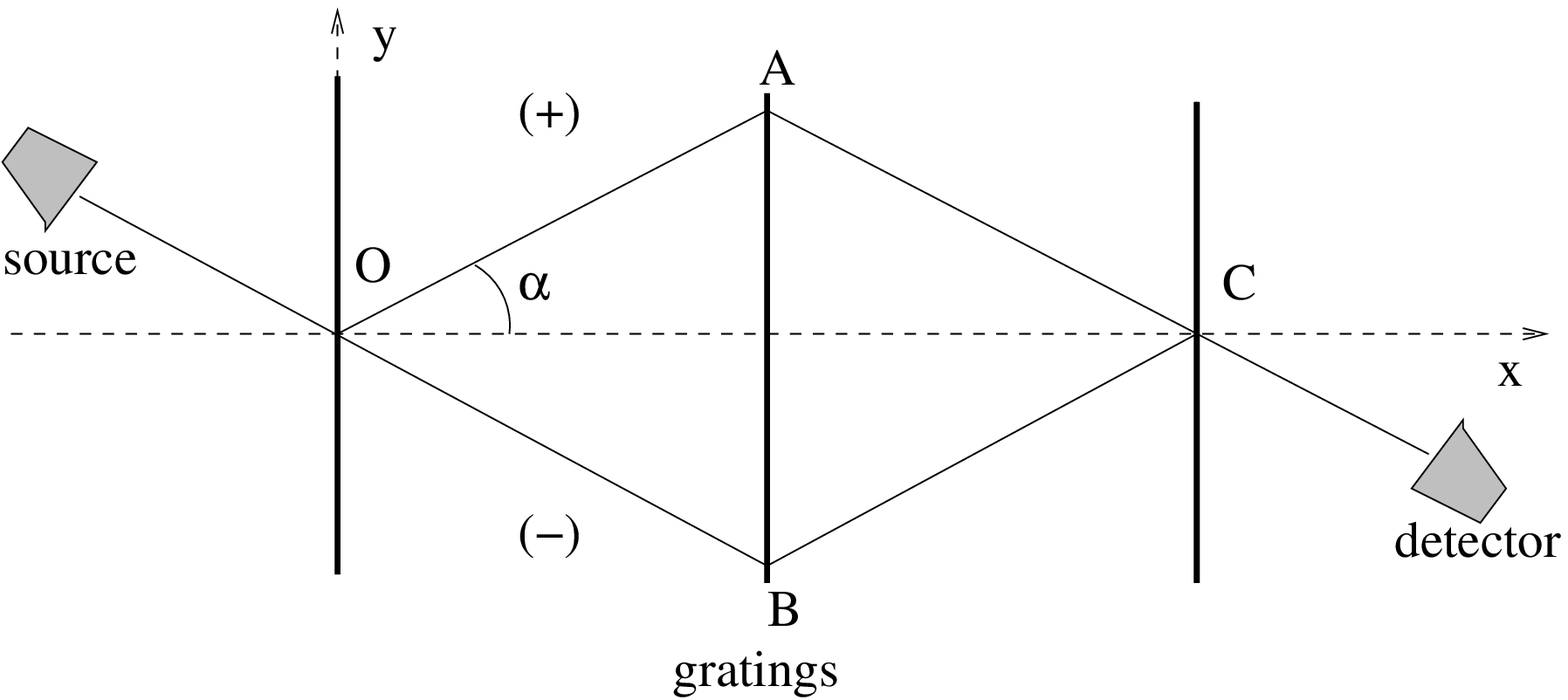}
\includegraphics[width=.3\linewidth]{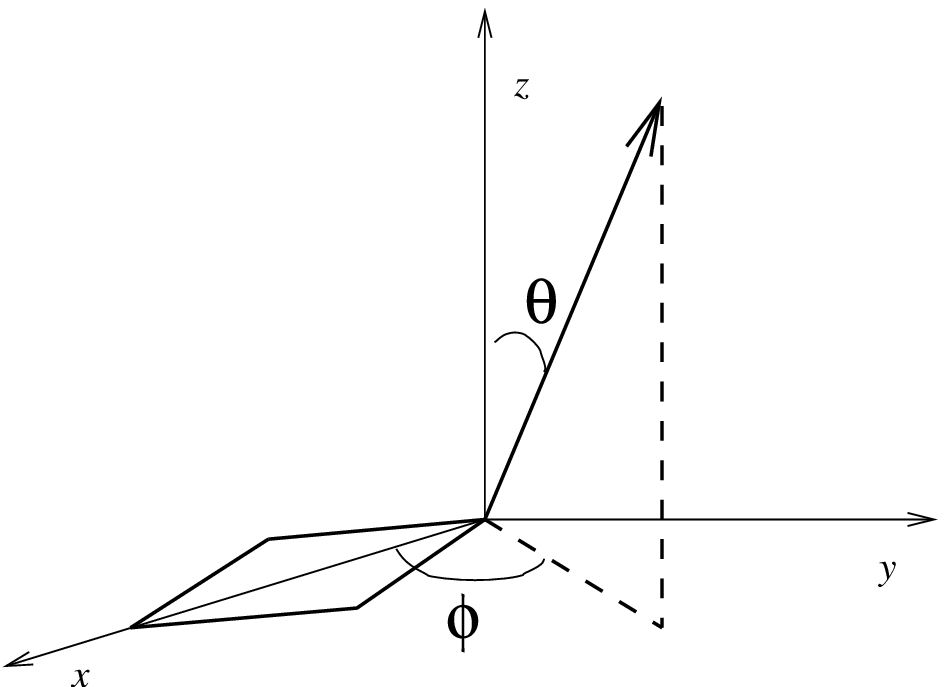} 
\centerline{\hspace{2cm}(a)\hspace{7cm}(b)}
\caption{Schematic configuration of the interferometer. Choice of the polar 
angles is also displayed.}\label{schema}
\end{figure}

The exact form of the response function depends on the details of the
experimental setup. For the sake of concreteness, we will consider the
Mach-Zender configuration which is displayed in Fig.\ref{schema}, which we
assume to be realized by means of properly chosen diffraction gratings.
Any other configuration and/or other deflecting device (mirrors, lasers) can be
studied along the lines which we are going to expose.
In the absence of GW, the classical upper and lower trajectories are
\be
X_{(\pm)}(\tau)= V_0 \tau c \alpha\,,\quad
\left\{\begin{array}{ll}\label{unpert}
Y_{(\pm)}(\tau)=\pm V_0 \tau s \alpha
& \quad {\rm for}\quad 0<\tau<T/2\\
Y_{(\pm)}(\tau)=\pm V_0 (T-\tau) s\alpha
& \quad {\rm for}\quad T/2<\tau<T\, ,
\end{array}\right.
\ee
with $c\alpha\equiv \cos\alpha$, $s\alpha\equiv \sin\alpha$ and we have 
from now on $t_i=0$, $t_f=T$.
We can now explicitly evaluate eq.~(\ref{phase3}) along each arm of the 
interferometer, that is from O to A, from A to C, do the same for the $(-)$ 
path and then take the difference.
With the help of eq.~(\ref{YofX}), the following expression is obtained:
\renewcommand{\arraystretch}{1.4}
\be\label{phazen}
\ba{cc}
\ds\frac{\hbar}{m}\Delta^{\pm}\Phi(C)=
\Delta^{\pm}\ds\int_0^{T}\left[-\frac{(V_0^y)^2}{V_0^x}w^x + V_0^y w^y -
\frac{1}{2}\dot h_{ij}V^i_0X_0^j(1-\hat n\cdot \vec{V}_0)\right]{\rm
  d}\tau+\\
\ds+\frac{V_0^2}{V_0^x}\Delta^{\pm}\pa{\Delta X_C-\Delta X_O}\, ,
\ea
\ee
\renewcommand{\arraystretch}{1.4}
where $\Delta^{\pm}$ means that the difference between the upper and the
lower path is taken. 
The first and the second terms in the integral and the last term are the ones 
discarded by Roura {\it et al.}, and in particular the parts containing 
$w^{x,y}$, which come from the expression of $\Delta Y$ through 
eq.~(\ref{YofX}), will give a contribution proportional to $\dot{h}$.

As to the last term, we observe that $\Delta^{\pm}(\Delta X_{C,O})=0$, that is 
the variation of the $x$-coordinate of the extrema of the trajectory are the 
same for the upper and the lower path and we can then discard the last term in 
eq.~(\ref{phazen}).

From now on we restrict to the case of normal incidence,
$n_i=\delta_{iz}$; in this case the symmetric geometry shown in
Fig.\ref{schema} makes the device sensible only to the
polarization represented by $h_{xy}\equiv h_{\times}$, because the $h_+$ 
polarization affects in the same way the $(+)$ and the $(-)$ paths.

The dependence of $w^{x,y}$ on $h_{\times}$ can be derived by
expanding to first order in $h$ the geodesic equation (\ref{geoeq})
around the unperturbed trajectories (\ref{unpert}).

In doing this one has also to take account of the junction conditions at
the grating. We model their effect as a rotation the atomic velocity by an 
angle $2 \alpha$\footnote{Strictly speaking, this relation is exact only up
  to terms of order $h(T/2)$. The derivation of the full relation is
  quite elaborate and it depends on the specific device deflecting the atoms. 
  However, since the correction to (\ref{rotate0}) does not depend on
  $\dot{h}$, which is the dominant contribution in the regime we are
  interested, we prefer to privilege clarity by presenting the
  simplified version of the junction conditions. Using other device than 
  a diffraction grating to deflect atom would result in different junction 
  condition, the modification affecting also the $\dot h(T/2)$ term.}:
\be\label{rotate0}
\vec{V}_{(\pm)}(T^+/2)-\vec{\cal V}_{A,B}={\cal R}[\mp 2\alpha]\cdot
\left(\vec{V}_{(\pm)}(T^-/2)-\vec{\cal V}_{A,B}\right)\,,
\ee
where we have denoted by $T^{+(-)}/2$ the instant of time immediately after 
(before) the interaction between the grating and the atoms.
The formula accounts for the velocities $\vec{\cal V}_{A,B}$ acquired
by the grating placed in A and B because of the passage of
the GW; these terms also contain the derivative of the GW and have
been inaccurately taken into account in \cite{Chiao:2003sa}.
Indeed we will see that their contribution suppresses the sensitivity in
the very high frequency regime.

By expanding (\ref{rotate0}) to first order in $h$ one obtains the
junction conditions for $\vec{w}$
\renewcommand{\arraystretch}{1.4}
\be\label{rotate1}
\begin{array}{l}
w^x_{(\pm)}(T^+/2)=c(2\alpha)\ w^x_{(\pm)}(T^-/2) +
(1-c(2\alpha)){\cal V}^x_{A,B}\pm
\left(w^y_{(\pm)}(T^-/2)-{\cal V}^y_{A,B}\right)s(2\alpha)\,,\\
w^y_{(\pm)}(T^+/2)=c(2\alpha)\ w^y_{(\pm)}(T^-/2) +
(1-c(2\alpha)){\cal V}^y_{A,B}\mp
\left(w^x_{(\pm)}(T^-/2)-{\cal V}^x_{A,B}\right)s(2\alpha)\,.\\
\end{array}
\ee
\renewcommand{\arraystretch}{1}
After some calculations, one arrives at the following expression
for the dephasing:
\renewcommand{\arraystretch}{1.4}
\be\label{time}
\frac{\hbar}{m}\Delta^{\pm}\Phi(C)&=&V_0 s\alpha\,{\rm tg}\alpha
\left\{\!2 V_0 s\alpha
\left(\int_0^{\frac T2}\!\!\!\!\!\!-\!\!\!\int_{\frac T2}^T\right)
\!h_{\times}(\tau){\rm d}\tau\!
\!+\!T\!\left[\frac {V_0 T}2
\dot{h}_{\times}\!\!\left(\!\frac T2\!\right)s\alpha
- \left({\cal V}^x_A-{\cal V}^x_B\right)\right]\right\}\, .
\ee
\renewcommand{\arraystretch}{1}

The square brackets in (\ref{time}) contain terms
involving, both explicitly and implicitly, the time derivative of the GW;
this is the most important contribution at frequencies $f$ such that
$T^{-1}\ll f$, and it is all due to the interaction with the central
grating, which introduces a discontinuity in $\vec w$ as an effect of
the junction conditions (\ref{rotate1}).
In particular, the term involving ${\cal V}^x_{A,B}$ represents the
direct effect of the state of motion of the grating, which imparts a kick to 
the atoms. We will show shortly that this effect cancels the preceeding 
$\dot h_{\times}(T/2)$ contribution in the limit $f \gg f_0$,
being $f_0$ the typical resonant frequency of the apparatus.
Such frequency is roughly given by $f_0=c_s/(2L)$, where $c_s$ is the sound 
speed of the material the experimental apparatus is made of, and $L$ its
linear size (of order $V_0 T/2$ in our case).

In order to obtain a closed form for the phase we need to express the 
velocities of the central grating with respect to the point O in terms
of known quantities. This can be done by solving their geodesic equations, 
which are damped harmonic oscillator equations with a source term given by the 
GW:
\be \label{eqmotoap}
\ddot {\cal X}_{A,B}^i +\frac{2\pi f_0}Q{\cal \dot{X}}_{A,B}^i+
(2\pi f_0)^2 {\cal X}_{A,B}^i =\frac 12 \ddot h^i_j {\cal X}_{A,B}^i\,,
\ee
where $Q$ is the quality factor of the oscillator and $\vec {\cal X}_{A,B}$
the position of the grating. Being 
\be
h(t)=\int {\rm d}f \, \tilde h(f) e^{-i2\pi f t}\,,\qquad
 \tilde h(f)=\int {\rm d}t\ h(t) e^{i2\pi ft}\,,
\ee
our convention for the Fourier transform, the perturbed velocities of the 
gratings can be easily written as 
\be \label{xsol}
\tl {\cal V}^x_{A}- \tl {\cal V}^x_{B}
\equiv- 2 \pi i f \left(\tl {\cal X}^x_{A}-{\cal X}^x_{B}\right)=
- 2 \pi i f \frac{\tilde{h}_{\times}}
{1-f_0^2/f^2+if_0/(fQ)} \frac{V_0 T}{2} s\alpha\,,
\ee
where in the Fourier representation of the
solution we neglected the transient term, i.~e. the motion of the
apparatus {\emph not} induced by $h$,
which is equivalent to assume that it was at rest before the arrival of the GW.

We can finally write the result for the Fourier transform of the dephasing. 
Here we report only the terms which dominate in the regime $T^{-1}\ll f$,
where the matter-wave interferometer is more performant: 
\renewcommand{\arraystretch}{1.4}
\be\label{fourier}
\displaystyle
\tilde \Delta^{(\pm)}\Phi(f)=\frac{m}{\hbar} A\ {\rm tg}^2 \alpha\ (- 2\pi if) 
\left(1-\frac{1}{1-f_0^2/f^2+if_0/(fQ)}\right) \tl h_{\times}\, ,
\ee
\renewcommand{\arraystretch}{1}
where $A=(V_0 T)^2\  s\alpha\ c\alpha/2$ is the area of the interferometer.

We now see explicitly that for $f/f_0\gg 1$ the $\dot h_\times$ effect 
vanishes. This was to be expected as in this limit both the atoms and the
diffraction gratings are freely falling and during the deflection of the
atoms by the experimental apparatus, assumed instantaneous,
the GW has no net effect. Typical time scale of such interactions
are smaller than the $\mu sec$, so indeed instantaneous in our problem, see 
Sect.\ref{noise}.
We underline that by using other configurations and/or
deflecting devices, a response function quantitatively different from 
(\ref{fourier}) can be obtained. For instance assuming that the atoms are
deflected by bouncing off mirrors (which turn out to be the most efficient 
devices, see sec.\ref{tech}) instead of by grating rotation at $A$ and 
$B$, the prefactor of $\dot h$ would change from $A\,tg^2\alpha$ to $A$, and 
dependence on $h(0)-h(T/2)$ might be introduced; however the response function 
will have the same qualitative properties, that is the presence of a term 
proportional to $f \tl h$ , which is however suppressed for $f \gg f_0$.

\subsection{Generic incidence direction}
For a GW impinging on the interferometer from a generic direction
\be
\hat\Omega\equiv(\sin\theta\cos\phi,\sin\theta\sin\phi,\cos\theta)
\ee
eq.~(\ref{fourier}) needs to be generalized.
The two possible polarizations can be written in terms
of two mutually orthogonal unit vectors $\hat{m}$ and $\hat{n}$ as
\be
\ba{l}
e^{+}_{ab}=\hat{m}_a \hat{m}_b-\hat{n}_a \hat{n}_b\, ,\\
e^{\times}_{ab}=\hat{m}_a\hat{n}_b+\hat{n}_a\hat{m}_b\, .
\ea
\ee
The unit vectors $\hat{m},\hat{n}$ must lie on the plane orthogonal
to $\hat\Omega$ and must be mutually orthogonal; one possible choice is
\be
\ba{l}
\hat{m}=(\cos\theta\cos\phi,\cos\theta\sin\phi,-\sin\theta)\, ,\\
\hat{n}=(-\sin\phi,\cos\phi,0)\, .
\ea
\ee
Keeping the axes as in fig.\ref{schema} the interferometer will still be 
sensitive to the $h_{xy}$ component of the GW, but for generic incident
direction $\hat\Omega$ one has $h_{xy}=F_+h_++F_\times h_\times$,
where the \emph{pattern functions} $F_{+,\times}$ are given by
\renewcommand{\arraystretch}{1.4}
\be\label{F+x}
\ba{l}
\displaystyle
F_+=\frac{1+\cos^2\theta}2\sin 2\phi\,,\\
F_\times=\cos 2\phi\cos\theta.
\ea
\ee
\renewcommand{\arraystretch}{1}
With respect to a different pair of axis $m',n'$, rotated with 
respect to $m,n$ by an angle $\psi$ in the $m,n$ plane (it is enough to have 
$0<\psi<\pi/4$), the $F_{+,\times}$ of (\ref{F+x}) becomes
$F'_+=\cos 2\psi F_++\sin 2\psi F_\times$, $F'_\times=\cos 2\psi F_\times-
\sin 2\psi F_+$.
The simple case reported in eq.~(\ref{fourier}) corresponds to the choice
$\theta=\phi=\psi=0$.\\
Strictly speaking this is not the only effect which has to be taken into
account to describe an oblique incoming direction, as
the action (\ref{action1}) involves a term proportional to $\hat n\cdot \vec
V_o$, which is not vanishing in the case of not normal incidence.
Anyway this effect represents a relativistic correction to the
leading term and thus it is negligible.

\section{Analysis of the noise}\label{noise}

Let us now concentrate again on the case of ideal incidence angle and 
polarization. The sensitivity of this kind of interferometer is determined by 
all the sources of noise that enter the actual measurement of $\Phi(t)$.

\subsection{Shot noise}
The possibility to detect GW's with an atomic
interferometer is fundamentally limited by phase/particle-number
uncertainty, and so by readout shot noise in the atomic detector.
In the case of atomic interferometry it is a fair assumption
that the total number of atoms detected per unit of time is perfectly known.
In that case \cite{Dowling:1993}, assuming contrast and detection efficiency
equal to 1 and a low density beam, 
the error on phase measurement is inversely proportional to
the square root of the number of detected atoms $N$
\begin{equation}
  \label{eq:phi-error}
  \sigma_{\Phi}^{\rm Shot} =\frac{1}{\sqrt N}=
\frac{1}{\sqrt{{\cal F} t_{\rm b}}}\, ,
\end{equation}
where ${\cal F}$ is the atomic flux and $t_{\rm b}$ the averaging time.
If the phase shift to be measured is time independent, in
principle $t_{\rm b}$ can be increased (collecting more atoms) to reduce the 
error on the phase.
However if the phase shift is time dependent, it is useful to perform a
discrete Fourier analysis. To detect a variable $\Phi(t)$ we have to bin
the data over a time interval $t_{\rm b}=\Delta t$ thus obtaining a row of data
$\Phi(t_j)$.
We assume that the
experiment is calibrated so that the number $N_j$ of atoms detected at each
bin is given by
\be \label{expsetup}
N_j={\cal F} \Delta t(1+\Phi(t_j))\, .
\ee
Then each of the $\Phi(t_j)$ is a random variable,
with average zero (in the absence of signal) and variance
$\sigma_j=1/\sqrt{{\cal F} \Delta t}$.
The Fourier transform of the data reads
\be
\tilde\Phi(f_k)=\Delta t\ \sum_{j=1}^M e^{-i2\pi kj/M}\Phi(t_j)\, ,
\ee
where the discrete frequencies are $f_k\equiv k/t_{\rm ex}$,
with $k$ in the range $1\leq k\leq M$, being $M$ the number of bins,
and we also defined the total experiment time $t_{\rm ex}=M\Delta t$.
Using the central limit theorem we can state that the $\tilde\Phi(f_k)$'s
are Gaussian distributed random variables with vanishing mean and
variance $\sigma_k$ given by the sum in quadrature, thus obtaining (in
the limit $M\gg 1$)
\be \label{eq:sigmak}
\sigma_k=\sqrt\frac{t_{ex}}{2\cal F}\,,
\ee
for both the real and the imaginary part of $\tilde\Phi(f_k)$.
We note that $\sigma_k$ grows with the integrated time of the
experiment (we are \emph{summing}, rather then averaging, when we take the
Fourier transform) and it is frequency independent. The latter property may 
seem
counterintuitive but one has to remember that even if higher frequencies
are characterized by higher relative errors on a range of time $t_k\equiv
1/f_k$, nevertheless $t_k$ fits more times in the total experiment time
and the two things compensate.
Notice that in the previous analysis of the shot noise, it was implicitly
assumed that the binned phase was constant over the binning time.
However the result is unchanged even if the previous assumption is relaxed
\cite{Dowling:1993}.

If we now plug eq.(\ref{eq:sigmak}) into the relation (\ref{fourier}),
to be more precise, between 
$\tl h(f)$ and $\tilde\Phi(f)$, we can estimate the minimum detectable
signal as
\be\label{hmin}
\tl h_{\rm min}(f)\simeq \frac{\hbar}{2 \sqrt{2} \pi}
\frac{\sqrt{t_{\rm ex}/{\cal F}}}{m A {\rm tg}^2 \alpha f|F_h|}\,,
\ee
where $F_h(f)$ can be derived from eq.(\ref{time}) 
\be
{\rm e}^{i \pi f T} F_h(f)=1-\frac{1}{1-f_0^2/f^2+if_0/(fQ)}-
\pa{\frac{\sin\pa{\pi fT/2}}{\pi fT/2}}^2\,.
\ee
We note that $F_h$ goes to zero in both the low and high frequency limits,
$fT\to 0$ and $f>f_0$, it is of order 1 for $1/T< f< f_0$ and
attains its maximum at resonance, where $F_h\sim Q$.

To make contact with common literature on GW detectors it is convenient
to express the sensitivity in terms of
$h_c(f)\equiv\tl h_{\rm min}(f)/\sqrt{t_{\rm ex}}$.
$h_c$ has dimension of square root of time and it is
measured in ${\rm Hz}^{-1/2}$; the main advantage of its definition is
that it does not depend on the observation time $t_{\rm ex}$.

\subsection{Other noise sources}
As previously stated,
the matter-wave interferometer reaches its best sensitivity at 
$f\simeq f_0=c_s/(V_0 T)=c_s/(2L)$, while for $f>f_0$ the sensitivity is 
drastically reduced because in this regime the central grating is not anymore 
rigidly attached to the beam splitter and is thus practically comoving with
the atoms.

Actually, the maximum detectable signal frequency $f_{\rm max}$ is
related to the binning time $\Delta t$ through $f_{\rm max}\simeq 1/\Delta
t$. Then, for a given $\Delta t$, we must ensure that the flux is
large enough in order to have enough atoms to arrive in each bin, that
is we must require ${\cal F}\gg f_0$.

Moreover, we remind that in the previous section
we neglected the spatial dependence modulation of the GW.
This can be done in the long wavelength approximation,
which ceases to be accurate when the wavelength of the
GW is of the same order of magnitude of the length of
the atomic paths. This enforces the condition
gives $f_0 L\ll 1$, being $L$ the total length of the interferometer;
this condition is always satisfied because of the smallness of the
sound speed $c_s$ with respect to the speed of light.

Even with these qualifications, beside the shot noise there are other possible 
noise sources that should be taken into account.

First of all, since the atomic beam is prepared with a finite velocity
spread $\delta v_0$,
then there is the possibility atoms entering the apparatus at the same
time arrive
at the detector at different time bins just because of this spread, thus
introducing spurious noise in the signal.
This leads to a minimum binning time $t_{\rm min}\simeq T \delta
V_0/V_0$ which thus enforces the condition (here and henceforth we
systematically neglect factors of order 1)
\begin{equation}\label{binning}
f_0<\frac{V_0}{T\delta V_0}\, .
\end{equation}

Moreover, we must take into account of the thermal noise affecting the 
deflecting devices, that is the influence of the thermal vibrations of the 
gratings (or mirrors) on the atoms' movement.

By a straightforward use of the the fluctuation dissipation theorem we can 
compute the (single sided) power spectrum of the velocity of the deflector 
$S_v$ as
\be
S_v(\omega)=4k_B T\,{\rm Re}\pa{\frac 1{Z(\omega)}}\,,
\ee
being $Z(\omega)$ the impedance function, i.e. the function relating the 
Fourier transform of the force and the velocity according to 
$\tilde F(\omega)=Z(\omega)\tilde v(\omega)$.
The explicit form of the impedance can be deduced from (\ref{eqmotoap}) and 
(\ref{xsol}) and it gives
\be \label{sv}
S_v=\frac{4k_BT}{M_{\rm def}}\frac{\omega^2\omega_0/Q}{\pa{\omega^2-\omega_0^2}^2+
\pa{\omega_0\omega/Q}^2}\,,
\ee
being $M_{\rm def}$ the deflector mass. 

This induces a noise power spectrum in the phase, which can be
estimated according to eq.~(\ref{time}) as
\be
S_{\Phi, v}=\left(\frac{m}{\hbar}V_0 T s\alpha {\rm tg}\alpha\right)^2
S_v\, .
\ee
To keep under control the thermal noise we must require that
$S_{\Phi, v}<S_{\Phi, shot}=1/2{\cal F}$;
given the shape of the spectra, it is enough to impose the constraint
at resonance, thus obtaining the following estimate:
\be\label{tmax}
T< 10 {\rm K}\pa{\frac{M_{\rm def}}{1{\rm Kg}}}\pa{\frac{Q}{10^5}}^{-1}
\pa{\frac{\cal F}{10^8{\rm s}^{-1}}}^{-1}
\pa{\frac{\omega_0}{4\cdot 10^5{\rm Hz}}}
\pa{\frac{L}{0.15 {\rm m}}}^{-2}\pa{\frac m{4m_p}}^{-2}\,,
\ee
which has been computed for $\alpha=\pi/4$.
This means that the apparatus can not work at room temperature; on the
other hand the required temperature is not very severe and can
be reached with standard cryogenic techniques.

Other kinds of thermal noise do not give more severe conditions than
(\ref{tmax}). The atomic-thermal shot noise (deformation of the
deflector because of the heating caused by the impinging beam) is completely
irrelevant here because of the low power of the atomic beam.

The Brownian motion of the deflector surfaces (structural damping) and the 
thermoelastic damping can be estimated along the line of 
\cite{Braginsky:1999rp}. The resulting induced spectral density of the
position of the deflector surface $S_x$ has to be compared with 
$S_v/\omega^2$ derivable from (\ref{sv}). With different numerical values for 
the thermo-mechanical specifications of the deflector (Young 
modulus, mass density, specific heat, expansion and diffusion coefficients)
and an atomic beam size of $1$mm we find that at resonance and at a
temperature of $10$K these noises are several orders of magnitude smaller than 
the noise induced by $S_v$. The noise induced by the structural damping is 
competitive with the noise induced by Brownian motion of the deflectors 
characterized by eq.(\ref{sv}) only well outside the resonance (at a frequency 
three order of magnitude smaller than $f_0$), where the interferometer 
sensitivity is anyway degraded by the peculiar behavior of the response 
function (\ref{hmin}).

Finally there are other effects, related specifically
to the details of the interferometric apparatus used, which determine
a reduction of the contrast of the interferometric figure and of the atomic 
detection efficiency. They correspond to the replacement of 
eq.~(\ref{expsetup}) with the expression 
$N_j=\eta {\cal F} \Delta t(1+\frac{\Phi(t_j)}{C})$ where $\eta$ is the 
detection efficiency and $C$ is the contrast. Consequently, we can take these 
two effects into account by multiplying eq.~(\ref{hmin}) by a factor $\eta/C$.

\subsection{Technical specifications and best sensitivity} \label{tech}

We now compute a realistic estimate of the sensitivity of an atom 
interferometer. Its size is a crucial factor, as the sensitivity
increases with the area but at the same time it increases with the signal
frequency and the maximal frequency $f_0$ is inversely proportional to the 
length of the arms $L$.\\
With $L=15$ cm and using the sound speed of graphite ($c_s=18350$m/s) we can
have $f_0\simeq 60$ kHz. It is also important that the opening angle of
the interferometer is not too small; such a parameter depends on the
technology employed.

If nanofabricated diffraction gratings are used one
has $\tan \alpha \sim 2 \hbar /m V_0 a$ with $a$ the grid
spacing. Typical values used in interferometers are $a \sim 200 nm$
giving $\tan \alpha \sim 10^{-4}$ for heavy atoms ($V_0\sim 10^3$ m/s, 
$m\sim 200 m_p$, being $m_p$ the proton mass). However gratings with smaller
spacing up to $a=32$nm have been built \cite{gratings}, as pointed out in
\cite{Chiao:2003sa}.

Reaching a large $\alpha$ with optical beam splitters is instead
challenging, since one has to rely either on very small wavelengths or
on high order Bragg scatterings. In the latter possibility, very well
collimated beams are required and in general intensity is strongly
reduced.

Using mirrors large angle deflection can be achieved allowing to build a 
squared interferometer. 
At present such mirrors have been built using crystal
surfaces \cite{helium1} and utilized with Helium atoms;
it follows that for our device it is advisable to use Helium atoms,
for which fluxes ${\cal F}\sim 10^8{\rm sec}^{-1}$ can be reached.
The choice of light atoms like Helium is also necessary to overcome crystal
surface imperfections \cite{helium2}.
Given the current ability on preparing crystal mirrors, it is unlikely
that heavy atoms like Cesium will be used with crystal mirrors in a near
future \cite{Chiao:2003sa}.
For helium we can take a value for the atoms speed of $10^3$ m/s, with 
$\delta V_0/V_0\sim 10^{-3}$, which ensure the validity of (\ref{binning}).

Finally, the sensitivity at the resonance can be improved by choosing
a high $Q$ for the oscillator: this means paying a price in terms of
bandwidth of the detector, but we will see that an high quality factor is
necessary for obtaining an acceptable sensitivity.
Actually matter interferometers are intrinsically wide band, but they can be 
turned into narrow band detectors in order to increase their peak 
sensitivity.

By building an interferometer with the previous specifications, one
can make it work efficiently at the resonance frequency $f_0\simeq 60$
kHz, where it attains its peak sensitivity:

\begin{eqnarray}\label{maxsen}
h_c^{min}&\simeq&\frac C{\pi\sqrt{2}\eta}\frac 1{\sqrt{\cal F}}
\frac{\hbar}{m}\frac {1}{L \frac{s^3\alpha}{c\alpha} Q c_s} \simeq
2\cdot 10^{-20}{\rm Hz}^{-1/2}\times\nonumber\\
&\times& \pa{\frac{C/\eta}{10}} \pa{\frac{\cal F}{10^8 {\rm s}^{-1}}}^{-1/2}
\pa{\frac m{4m_p}}^{-1}\pa{\frac L{0.15{\rm m}}}^{-1}
\pa{\frac Q{10^5}}^{-1}\pa{\frac{c_s}{1.8\cdot 10^4\rm{m/s}}}^{-1},
\end{eqnarray}
for $\alpha=\pi/4$ and $m_p$ the proton mass.

\begin{figure}
\centering
\includegraphics[width=.9\linewidth,angle=0]{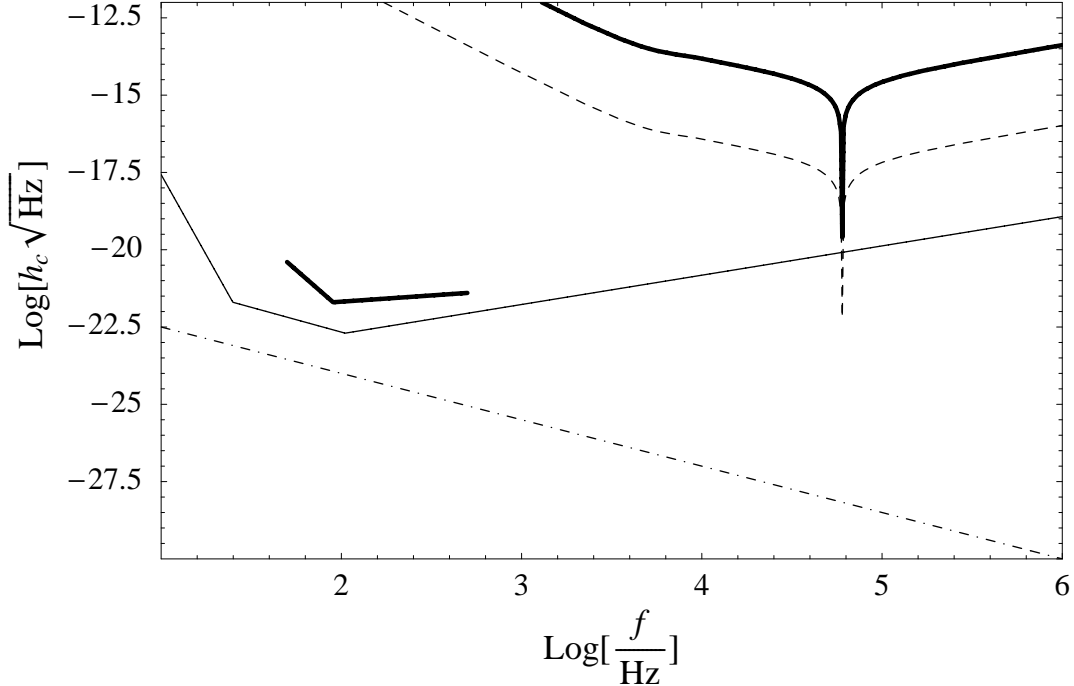}
\caption{
  Thick solid lines: sensitivity curve for a Helium interferometer with 
  $L=0.15$m, $\alpha=45^o$, $f_0=60$kHz, $Q=10^5$, 
  ${\cal F}=10^8{\rm atoms}/{\rm s}$, total efficiency of 10\%, 
  and sensitivity curve for the S3 run of LIGO \cite{Abbott:2005ez}.
  Solid line: sensitivity goal for LIGO1 extrapolated to the MHz.
  Dashed line: sensitivity curve for two correlated matter interferometers.  
  Dash-dotted line: BBN bound on a stochastic background of GW.}\label{sensit}
\end{figure}

This sensitivity can be compared with the one of a light interferometer at the 
same frequency (at the frequencies we are interested in, $h_c^{\rm LIGO}$ is 
dominated by the shot noise). In Fig.\ref{sensit} we compare the LIGO 
sensitivity curve, as it has been recently reached \cite{Abbott:2005ez} 
in the S3 run, with the sensitivity curve for a matter-wave interferometer 
endowed with $L=15cm$, ${\cal F}=10^8{\rm atoms}/{\rm s}$, $\tan \alpha = 1$, 
and a total efficiency of 10\%. 
An extrapolation to high frequencies of the LIGO1 sensitivity goal is also 
showed.

\section{Stochastic signal: correlation of two interferometers} \label{corr}

For a given $h_c$ one needs a much different gravitational
wave amplitude according to whether the signal to be detected is a burst or a
monochromatic wave.
Let us consider e.g. a burst of amplitude $h_B$ and duration time
$\tau_B$, its Fourier transform is $\tl h_B\simeq h_B \tau_B$
thus giving $h_c\simeq h_B \sqrt{\tau_B}$. So, in order to be detectable
by an interferometer with sensitivity $h_c$ a burst must have an amplitude 
such that
\be
h_B>\frac{h_c}{\sqrt{\tau_B}}\, .
\ee
This has to be compared to a periodic signal, say a
cosine of frequency $f_P$ and amplitude $h_P$. The Fourier transform
in this case gives $\tl h(f_P)\simeq h_P t_{\rm ex}$, thus the signal is
detectable if
\be
h_P>\frac{h_c}{\sqrt{t_{\rm ex}}}\, ,
\ee
and since $t_{\rm ex}\gg \tau_B$, one has that periodic signals are
easier to detect than bursts.

Actually the dependence of the minimum attainable $h$, for a given $h_c$, on 
the nature of the signal is even stronger if correlation of detectors is taken
into account.

In the case of a GW of stochastic origin a signal even lower than the noise 
level could be detected, provided one has two detectors and their outputs are 
correlated.
We report here only the relevant formulas, addressing to \cite{Maggiore:1999vm}
for a detailed and systematic treatment.
By analyzing the output of two detectors for an observation time
$t_{\rm ex}$, and assuming that both the stochastic signal and the
detector sensitivity are slowly varying on a frequency range
$\Delta f$, one can reduce the minimum detectable amplitude according
to the following expression
\be
^{2d}h_c \simeq \frac{^{1d}h_c}{(t_{\rm ex}\Delta f)^{1/4}}\, ,
\ee
which, for $t_{\rm ex}=1$year and $\Delta f\simeq 1$Hz, gives an 
amplification factor of about $10^2$. 
Correlation of two light interferometers cannot be realized at high 
frequencies ($f\gtrsim$ 50Hz), as the maximum frequency of a signal which can 
be possibly correlated is given by the inverse distance of the two detectors, 
$f_0=6\cdot 10^4$Hz corresponding to a separation of roughly $1$Km. Of 
course correlation of a light interferometers with other kinds of devices is 
possible.

Given the moderate cost of a matter-wave interferometer, the construction of 
several matter-wave interferometers is an affordable task: the correlation of 
$n$ such detectors would increase the sensitivity to a stochastic signal by a 
factor $\sim n^{1/2}$ \cite{Christensen:wi} with respect to the $n=2$ case.

In the $50$ kHz region the only known candidate for
such a signal is a background of cosmological origin, whose strength
is limited by the so-called nucleosynthesis bound. The amount of gravitational
waves produced in the early Universe is constrained by the fact that it can
not carry too much energy density, as otherwise it will modify the production  
of elements during nucleosynthesis, which is predicted by the standard theory 
with a remarkable accordance with observations.

This gives a bound on the fraction on energy density that can be
carried by these relic cosmological gravitons:
\be
h_0^2 \Omega_{\rm gw}\leq 10^{-6}\, ,
\ee
where $h_0$ is a number of order one that parameterizes our ignorance about
the exact value of the Hubble parameter.
Since the energy density fraction in GW's and $h_c$ are related by
\be\label{omh}
h_0^2 \Omega_{\rm gw}\simeq
\left(\frac{f}{10^4 {\rm Hz}}\right)^3
\left(\frac{h_c}{10^{-24} {\rm Hz}^{-1/2}}\right)^2\, ,
\ee
one has that such a cosmological background is out of the reach even
of the combined operation of two small matter wave interferometers.
As this bound on $h_c$ goes as $f^{-3/2}$ at high frequency, it is
useless to push the matter-wave interferometer at extremely high frequency in 
the search for a background of gravitational origin.

There is a qualification to be done on the last statements:
in deriving eq.~(\ref{omh}) it is implicitly assumed that the spectrum
of the stochastic signal is smooth enough; the bound can thus be
avoided if a narrow peak is present in the spectrum.

\section{Conclusions and remarks}

We have studied the feasibility and the performance of a table-size
matter-wave interferometer for the detection of gravitational waves.

We find that its sensitivity in the high frequency regime is proportional to 
the time derivative of the gravitational wave, contrary to what is claimed in 
\cite{Roura:2004se}, but still this is not enough to compete with the large
light interferometers in their frequency bandwidth. However matter-wave 
interferometers can complement their light counterpart in the high frequency 
region (above $10$ kHz).

We have thus studied in detail the sensitivity of a matter wave interferometer 
around its maximal working frequency, and we have found that the
atomic beam quality and opening angle are the key factors for its 
sensitivity. Increasing the size improves its sensitivity by an area factor, 
but it decreases the frequency of highest sensitivity.
Peak sensitivity can be improved exploiting a resonance mechanism, thus 
making matter interferometers narrow band, but it is not excluded to use them 
as wide band detectors once technological advances (atom lasers for 
instance) will allow to lower noises.

The correlation of two or more matter-wave interferometers can improve the 
sensitivity for a stochastic signal, but the nucleosynthesis bound makes the 
detection of a cosmological background still out of reach.

In the frequency region above $10$ kHz the only astrophysical sources could be 
coalescing primordial black holes, whose mass can be smaller than the 
Chandrasekar limit by orders of magnitude. However the strength of the 
gravitational radiation they will eventually emit is too small in this 
frequency range and the signal is decreasing with the wave frequency and 
increasing with the mass of the coalescing objects, thus 
being stronger in the LIGO band rather than in the matter interferometer one.

\section*{Acknowledgements}

We are indebted to Michele Maggiore for raising our interest on the subject.
We thank Florian Dubath, Iacopo Carusotto and Gabriele Ferrari for useful 
discussions. 
The work of A.G. and R.S. is supported by the Fonds National Suisse.
M.P. wishes to thank the
theoretical physics department of the University of Geneva for kind
hospitality during the completion of this work.

\end{document}